\documentclass[onecolumn,prd,amsmath,amssymb,floatfix]{revtex4}

\usepackage{amsmath,amssymb}
\usepackage{bm}
\usepackage{epsfig}

\newcommand{\bpsi}{\bar{\psi}}

\usepackage{slashed}

\begin{document}

\title{ Vacuum polarization 
corrections to  low energy quark effective couplings
}

\author{Ademar Paulo Jr.$^{1,2}$ and Fabio L. Braghin$^1$}

\affiliation{$^1$ Instituto de F\'\i sica, Univ. Fed. de Goi\'as, P.B.131, Campus II, 
74001-970, Goi\^ania, GO, Brazil
\\
$^2$ Inst. Fed. de Educ., Ci\^encia e Tec. do Tocantins, Campus Palmas, 77021-090, Palmas, TO, Brazil}

\begin{abstract}
In this work  corrections to  low energy punctual effective quark couplings 
up to the eighth order are calculated by considering  vacuum  polarization effects with  
the scalar quark-antiquark condensate.
The departing point is  a   QCD-based NJL model.
By separating the quark field into two components,
one that condenses and another one for interacting  quarks, the former  is integrated out
with the help of usual auxiliary fields
and an effective action in terms for interacting  quark fields   is found.
The scalar  auxiliary field reduces to the quark-antiquark condensate  in the 
vacuum  and the  determinant is expanded in powers of the 
quark-antiquark bilinears generating 
  chiral invariant effective 2N-quark interactions ($N=2, 3...$).
The corresponding coupling constants and  effective  masses 
are estimated, and the general trend is that for  increasing  the effective gluon mass the values of 
the effective coupling constants decrease. All  the values are in good agreement with phenomenological fits.
\end{abstract}


\maketitle

\section{Introduction 
and  extended Nambu-Jona-Lasinio model from QCD}

In spite of spectacular progresses in  lattice calculations
 it still is extremely important to 
rely the description of hadron, and more generally nuclear,
 processes on QCD based  hadron effective models.
These models  are expected to incorporate the most important 
symmetries and properties of the fundamental theory (QCD)
such as the chiral symmetry and its spontaneous breakdown.
Many  effective models have been proposed to describe the low energy regime of the 
 QCD phase diagram being 
the quark  Nambu-Jona-Lasinio (NJL) model \cite{NJL,NJL2,NJL3,NJL4}
one of the most emblematic QCD effective  models 
due to its relative simplicity with the corresponding power  to describe some aspects of low energy
 hadron 
physics due to
 the spontaneous chiral symmetry breakdown  by means of the 
chiral  condensate  $\langle \bar{\psi} \psi \rangle$.
In spite of the recent controversy on the formation of the scalar quark condensate
 \cite{in-hadron-1,in-hadron-2,reinhardt-weigel}
it is widely recognized  its contribution  for example for 
the nucleon  and quark  masses \cite{ioffe-ppnp,qqbar,qqbar2},
even if gluon dressing might  also be important \cite{lavelle-mcm}.
It is however very important to extend its validity and refine its predictions by including
further effective quark interactions \cite{enjl-1,e-njl2,andrionov,8th,stability-8th,stability-8th-2,polyakovloop-njl,C-R}. 
For the  high energy limit of the phase diagram, 
Polyakov loops  were
included to incorporate the deconfinement phase transition \cite{polyakovloop-njl},
extending the validity of the model.
In spite of being suitable for describing low energy physics, this punctual effective interaction 
has also been envisaged for high energies \cite{LHC1}.
Osipov {\it et al}  have found
that an eighth order quark interaction term restores the stability of the vacuum 
\cite{stability-8th,stability-8th-2}
that is unfavored by the  sixth order SU(3) 't Hooft 
  interaction
 \cite{instability}. 
Whereas the 't Hooft  interaction was found long time ago from QCD grounds \cite{thooft}
 the eighth order term has already 
 received different approaches  which do not necessarily
exclude each other  \cite{simonov,simonov-2,shuryak,barros-braghin}.
Since one can think about including progressively higher order effective quark interactions,
it is interesting to note  that  
 the longstanding problem of the convergence of the QCD 
action in powers of quark currents  \cite{weinberg,qcd-expansion,scherer-chpt}
might also receive some insight from 
the microscopic calculations of multiquark interactions.
Of course eventually one might have  to  avoid
double counting effects, what might be extremely difficult to assess.
Although the emergence of such higher order   interactions has already  been  addressed
in the last decades, their contribution 
to the structure and reactions of the new (heavier)  hadrons that were 
proposed to have multiquark structure, see for example Refs.   \cite{tetraq},
 is not really understood.
In this work we derive low energy effective quark couplings 
due to the vacuum polarization with the chiral condensate.

A general quark effective action
obtained by integrating out  gluons from the QCD action
 can given by
\cite{ERV,C-R,barros-braghin}:
\begin{eqnarray} \label{Seff}  
S_{eff}  [\bar{\psi}, \psi] = \int_x \left[
\bar{\psi} \left( i \slashed{\partial} 
- M \right) \psi 
- 
 \frac{1}{2}\int_y j_{\mu}^b (x) 
(R^{\mu \nu}_{bc} )^{-1} (x-y) j_{\nu}^{c} (y) 
 \right]
 + S_A ,
\end{eqnarray}
Where the color  quark current is $j^{\mu}_a = \bar{\psi} \lambda_a \gamma^{\mu} \psi$,
 the sums in color, flavor and Dirac indices are implicit,
the kernel $R^{\mu \nu}_{bc}$ is the gluon propagator that might depend on 
auxiliary variables
and the last  term $S_A$   
 corresponds to  terms due  the gluon integration, 
including the  gluon determinant and  ghost integration if needed,  
 and eventually with  dependence  on auxiliary variables \cite{barros-braghin,DVBG,VKvAV,dudaletal-JHEP}.
To investigate the flavor structure of the model, one might  perform a Fierz transformation 
in the current-current interaction from which
a NJL emerges.
Several QCD condensates have been proposed besides the quark-antiquark, and
two gluon condensates have gain further attention:
the order 2 condensate  ($\langle A^2 \rangle$) 
and the order 4 condensate ($\langle F^{\mu\nu}F_{\mu\nu} \rangle$).
The interplay of the second gluon condensate with  quark effective interactions
was already considered for example  in Ref. \cite{E-V-91-conden}.
We wish to consider  the former since it 
has been related to a possible effective  gluon mass \cite{cornwall,gluonmass,gluonmass-1}
that has been seemingly found in several other analytical calculations 
 \cite{li-shakin,RGZ,GZ-gluonmass,GZ-gluonmass-1,serreau-tissier,strauss-etal,gluonpro,ver-ver11}
and in numerical and  lattice calculations  
\cite{cucchieri-2011,lattice,lattice-2,gluonmass,gluonmass-1,njl-gluon-propag,cristina}.
Our departing point  therefore 
is the NJL  of  Ref. \cite{barros-braghin} with a gluon condensate of order 2, 
although the
NJL model can be obtained   from QCD within different considerations
\cite{qcd-njl,qcd-njl-2}.
A   different approach was considered by Simonov within
the instanton gas model to derive effective quark interactions \cite{simonov,simonov-2,shuryak}
and we will not investigate to what extent these two approaches provide 
sort of double counting  effects or not.

The work is organized as follows.
In Section II the  Nambu Jona Lasinio model  induced by 
 the gluon condensate of order 2 ($\langle A_{\mu} A^{\mu} \rangle \sim \phi_0$)
is considered such that 
 the quark content   is separated 
into  two components:  the quasi-particle sector corresponding to
the interacting quarks and the one corresponding to the
condensed quarks, such that 
$\overline{\psi}\psi \to  (\overline{\psi} \psi)_c + \overline{\psi} \psi$,
preserving explicitly  chiral symmetry.
The variables $(\overline{\psi} \psi)_c$ are integrated out by introducing 
a set of usual  auxiliary variables $S_i, P_i$ that yields the 
scalar quark-antiquark condensate and a pseudoscalar variable.
 The (coupled) GAP equations of the auxiliary variables ($S_{i,0}, P_{i,0}, \phi_0$) 
are derived and solved in terms of an unique Euclidean covariant  cutoff yielding results 
in perfect agreement with well known effective masses from lattice and phenomenology.
In  Section III  the  quark determinant  is expanded in powers of the quark field, or quark flavor currents,
yielding   polynomial effective quark  interactions  whose couplings
depend on the two condensates $\langle \overline{\psi} \psi \rangle$ and $\langle A^2 \rangle$.
The values of the effective coupling constants are estimated using the 
same  cutoff, or conversely the  same gluon effective mass,  as the one considered for the GAP equations,
yielding values  comparable to those 
used in phenomenological fits in the literature.
In the last section there is a summary.

\section{The NJL and the scalar quark-antiquark condensate}

The  generating functional of  
the local  $SU (3)$ (flavor)  NJL  limit of the action (\ref{Seff})  
is given by \cite{NJL,qcd-njl,barros-braghin}:
\begin{eqnarray} \label{GF-NJL}
Z[\bar{\eta},\eta] = \int {\cal D} [\overline\psi, \psi]
exp \left\{  i \; \left[ S_{NJL} [\bar{\psi}, \psi] + \int_x \left( \overline{\eta} \psi + \eta \overline{\psi} \right) \right]
\right\}
\end{eqnarray}
where
\begin{eqnarray} \label{S-NJL}
S_{NJL} [\bar{\psi}, \psi] = \int_x \left\{
\bar{\psi} \left( i \slashed{\partial} 
- M \right) \psi 
+ 
\frac{g_4}{2} \left[
(\overline{\psi} \lambda_i \psi)^2 
+ (\overline{\psi} \gamma_5 \lambda_i \psi)^2
\right]  \right\} 
+ S_{A}
\end{eqnarray}
for $i = 0, ..., N_f^2-1$.
 For the  gluon determinant,  $S_A = - \frac{i}{2} \mbox{Tr} \log (R^{\mu \nu}_{ab})$, 
where Tr stands for sum over all indices and spatial integration.
The following truncated gluon propagator will  be considered:
\begin{eqnarray} \label{gluon-prop-phi}
(R^{\mu \nu}_{bc} )^{-1}(x-y)  =  \delta_{bc} \left[
(\partial^2  + c \phi(x) ) 
\left( g_{\mu \nu}
- \frac{\partial_{\mu} \partial_{\nu}}{\partial^2}
\right) + 
\frac{1}{\lambda} \partial_{\mu} \partial_{\nu} \right]^{-1} \delta^4(x-y),
\end{eqnarray}
where $\lambda$ is the  (covariant) gauge fixing  parameter.
The  transverse effective  gluon mass is therefore
 $M_G^{2} = c \phi_0$, being $\phi_0$ an auxilary variable for the gluon condensate 
$<A^2>$ , 
$g_4$  has dimension $({mass})^{-2}$ and it is of the order 
of $N_c^{-1}$,  at least in the leading order, and it is 
given by:
\begin{eqnarray} \label{g4}
g_4 = \frac{\beta}{M_G^2},
\end{eqnarray}
 where  $\beta$ a numerical factor accounting for the 
color and flavor structure of the model,
for the $SU(3)$ 
 $\beta =\frac{g^2}{9}$ where $g$ is the zero momentum QCD running coupling constant \cite{lattice}.

Let the quark field bilinears to be separated   into two components: 
the one  corresponding to the   quarks that condense ($(\overline{\psi} \psi)_c$)
and the other  to the  interacting quasi-particle  quarks,
analogously to  other formalisms, see for example in Refs.
\cite{variational,rg}.
This is done by considering that each quark bilinear, as well as the functional measure of 
the generating functional,  will be written as:
\begin{eqnarray}
\overline{\psi} \psi  \to (\overline{\psi} \psi)_c + \overline{\psi} \psi.
\end{eqnarray}
 This way chiral symmetry will  not explicitly broken.
A  further analysis within a renormalization group approach is outside the scope 
of this work.
Since $tr \left[ \gamma_5 \right] = 0$ and 
then $\langle \overline{\psi} \gamma_5 \lambda_i \psi \rangle = 0$,
 only even powers of the pseudoscalar bilinear will contribute.
The resulting  interaction term  can be written as:
${\cal L}_I 
= {\cal L}_q +{\cal L}_c +{\cal I}_{int}$ where:
\begin{eqnarray} \label{L-separated}
{\cal L}_q &=& g_4
\left[ (\overline{\psi} \lambda_i \psi)^2 +  (\overline{\psi} i \gamma_5 
\lambda_i \psi)^2 \right] ,
\nonumber
\\
{\cal L}_c &=& g_4
\left[ (\overline{\psi} \lambda_i \psi)^2_c +  (\overline{\psi} i \gamma_5 
\lambda_i \psi)^2_c \right] ,
\nonumber
\\
{\cal I}_{int} &=&  g_4
\left[  \overline{\psi} \lambda_i \psi \cdot (\overline{\psi} \lambda_i \psi)_c
+  \overline{\psi} i \gamma_5 \lambda_i \psi \cdot (\overline{\psi} i \gamma_5 \lambda_i \psi)_c
+  (\overline{\psi} \lambda_i \psi)_c \cdot  \overline{\psi} \lambda_i \psi
+  (\overline{\psi} i \gamma_5 \lambda_i \psi)_c \cdot  \overline{\psi} i \gamma_5 \lambda_i \psi \right].
\end{eqnarray}

The quark component  $(\bar{\psi} \psi)_c$  can be integrated out 
by introducing usual $SU(3)$ auxiliary variables $S_a, P_a$.
For that, the above  generating functional
is multiplied by the following unity integrals:
\begin{eqnarray} \label{aux-variab-1}
1 &=&  N' \; \int \; D [S_i, P_i ]
\; 
exp \;  \left[ - \frac{ i}{2  c_s} \int_x \;  \left[ S_i^2 + P_i^2
\right] \right]
,
\end{eqnarray}
where $c_s$ is a constant to be fixed and $N'$ 
a   normalization constant.
The fourth order quark  interaction   ${\cal L}_c$ cancels out
if 
 the following variable shifts with corresponding unit Jacobian are 
done:
$S^i \to S^i + 2g_4 (\bar{\psi} \lambda^i \psi)_c$ and 
$P^i \to P^i + 2g_4 (\bar{\psi}  i\gamma_5 \lambda^i \psi)_c$,
where we consider $c_s = 2g_4$.
One is left with the following linearized action for the component $(\bar{\psi} \psi)_c$
in terms of the auxiliary variables:
\begin{eqnarray} \label{L-q-1}
S_{NJL} & \to & 
\int_x \left[  \int_y \overline{\psi}_c \left( S^{-1} (x-y)  \right)  \psi_c
 - \frac{1}{4 g_4} 
\left( S_i^2 + P_i^2 \right) 
+ \bar{\psi}\left( i \slashed{\partial} - M \right)\psi + 
g_4
\left[ (\overline{\psi} \lambda_i \psi)^2 +  (\overline{\psi} i \gamma_5 
\lambda_i \psi)^2 \right]   
\right]
+ S_A,
\end{eqnarray}
In this equation:
\begin{eqnarray}
S^{-1} (x-y) = \left[ i \slashed{\partial} - M^* + 
2 g_4 \; \left( \; \lambda_i \; (\overline{\psi} \lambda_i \psi) + \lambda_i i \gamma_5 \;
( \overline{\psi} i \gamma_5 \lambda_i \psi) \; \right) \; \right] \delta^4(x-y)
\end{eqnarray}
being that the effective mass (matrix),
already 
receives the contribution from the auxiliary variables $S_i$ that 
will not vanish in the vacuum, i.e.:
\begin{eqnarray} \label{Meff}
M^* = M + S_i \lambda_i.
\end{eqnarray}

By integrating out the component $(\overline{\psi}\psi)_c$,
the resulting effective action for the quasiparticle quarks
with the  auxiliary variables
reads:
\begin{eqnarray} \label{S_eff-log}
S_{eff} & = &  
 i \; Tr \ln \left[ - i S^{-1} (x-y)\right] +
\int_x \left[ - \frac{1}{4 g_4} 
\left( S_i^2 + P_i^2 \right)
+ \bar{\psi}\left( i\gamma_{\mu}\partial^{\mu} - M^* \right)\psi +
g_4 
\left[ (\overline{\psi} \lambda_i \psi)^2 + (\overline{\psi} i \gamma_5 
\lambda_i \psi)^2 \right]  \right]
+ S_A.
\end{eqnarray}
Before expanding this expression in terms of the quark bilinears 
it is desirable to derive a set of GAP equations to determine the ground state
by extremizing this effective action with respect to the auxiliary variables, $\phi_0, S_i, P_i$.

\subsection{ Ground state: gluon condensate of order 2  and  the chiral scalar quark-antiquark condensate}

The gluon sector  of the  effective action (\ref{S-NJL}) 
will be replaced by:
\begin{eqnarray} \label{S-NJL-phi}
S_A= - \int_x \left[
 \frac{c }{4}\phi^2  \right] -  \frac{i}{2} Tr \ln \left[ 
R^{\mu \nu}_{bc} (x-y)
\right]  , 
\end{eqnarray}
so that effective action (\ref{S_eff-log}) can be written as:
\begin{eqnarray} \label{S_eff-full}
S_{eff} & = &
i \; Tr \ln \left[ S^{-1} (x-y) \right]  -  \frac{i}{2} Tr \ln \left[ 
R^{\mu \nu}_{bc} (x-y)
\right]  
\nonumber
\\
& &  +
\int_x \left[ - \frac{1}{4 g_4} 
\left( S_i^2 + P_i^2 \right)
-
\frac{c}{2} \phi^2 + 
\bar{\psi}\left( i\gamma_{\mu}\partial^{\mu} - M \right)\psi +
g_4 
\left[ (\overline{\psi} \lambda_i \psi)^2 + (\overline{\psi} i \gamma_5 
\lambda_i \psi)^2 \right]  \right] 
\end{eqnarray}

In the ground state  quasiparticle fields are  zero
 $\overline{\psi}, \psi \to 0$ and the effective potential for the 
gluon and quark-antiquark condensates ($\phi_0, S_i$, besides the pseudoscalar variable $P_i$) can be  
 extremized.
These equations are the following:
\begin{eqnarray} \label{extremizing}
\left. \frac{\partial {S}_{eff} }{\partial \phi} 
\right|_{\phi = \phi_0 , S^i = S_0^i}  &=&       0 ,
\nonumber
\\
\left. \frac{\partial { S}_{eff} }{\partial S_i} 
\right|_{S^i = S_0^i, \phi = \phi_0}  &=&          0,
\nonumber
\\
\left. \frac{\partial { S}_{eff} }{\partial P_i} 
\right|_{S^i = S_0^i, \phi = \phi_0}  &=&           0,
\end{eqnarray}
being that in all these equations $P_0^i=0$   is a trivial  and necessary  solution.
These 
 GAP equations, 
 in the Euclidean momentum space,  will be regularized by a covariant cutoff
yielding:
\begin{eqnarray} \label{gaps-phi-sigma}
 \phi_0  &=&   3 (N^2_c - 1) 
 \int \frac{d^4 k_E}{(2 \pi)^4} 
\left( \frac{1}{k^2_E +  M_G^2 }\right) + 
\frac{9 }{4 g^2 } \left( S_u^2 + S_d^2 +  S_s^2 \right),
\nonumber
\\
S_f^0 &=& 
\frac{  16 g^2 N_c  }{ 9 c \phi_0} \int \frac{d^4 k_E}{(2 \pi)^4} 
\left( \frac{M^*_f}{k^2_E + {M^*_f}^2 } \right)
,
\nonumber
\\
P_f^0 &=&  0 ,
\end{eqnarray}
where \textit{f=u,d,s} stands for up, down, strange  quarks
and the effective mass  can be written for each quark flavor as $M_f^* = M_f + S_f^{0}$.

In these equations,  the only free parameters are 
   the QCD coupling constant, which will be  given by
the zero momentum running coupling    $\alpha_s \equiv \frac{g^2}{4 \pi} = \frac{8.92}{N_c}$  
 \cite{lattice}, 
and the current quark masses, $m_u = 3$ MeV, $m_d = 6$ MeV and $m_s = 91$ MeV \cite{PDG}.
With these four  Lagrangian parameters  there is one covariant cutoff that solves the four 
gap equations,
fixing $N_c=3$.
Solutions for  typical values of the  effective gluon mass, found in
 lattice calculations,are shown in Table I
with the resulting effective quark masses.
For example, with   $M_G \simeq 650$ MeV 
 \cite{lattice,njl-gluon-propag,gluonmass,li-shakin} 
it   yields  $M^*_u= 297$ MeV, $M^*_d= 303$ MeV  and $M^*_s= 419$ MeV,
 for $\Lambda = 706$ MeV.
As expected this cutoff is  the usual NJL cutoff \cite{NJL2,NJL3,NJL4}
and certainly  higher than the $\Lambda_{QCD}$ 
\cite{lattice,lattice-2,cucchieri-2011}.
In Ref. \cite{E-V-91-conden} a similar model was developed by considering the order four gluon condensate
 ($\langle F_{\mu \nu}^2 \rangle$) with similar reasonably good results.

\section{ Effective quark interactions and coupling constants}

In this section  the  effective quark model  (\ref{S_eff-log}) will be
expanded in the   lowest order derivative expansion  
\cite{chan} 
 in powers  of  bilinears  $\bar{\psi} \Gamma  \psi$   (where $\Gamma$ is any 
combination of flavor $\lambda_i$ and $\gamma_{5}$) and the gluon sector
 $S_A$ of the model plays no role from here on.
If in one hand  an expansion of this kind  might impose
certain limitations on the resulting values of the effective couplings
since it is a perturbative treatment, on the other hand 
 it might require  instead a weak  strength of the quark fields
to assure its  validity.
Nevertheless, it might be a safe starting point
for investigating higher order quark effective interactions.
The expansion has the following shape:
\begin{eqnarray} \label{deriv-exp}
S_{eff} &\simeq& 
S_{eff,(0)} [\phi_0,S_i,P_i] 
 + \frac{1}{1! 1!}
 \int_{x_1, x_2}   \left.
\frac{\delta^2 S_{eff}}{\delta \overline\psi (x_1) \delta  \psi(x_2)} 
 \right|_{\psi=\bar{\psi}=0}
 \overline\psi (x_1) \psi (x_2)
\nonumber
\\
&&
+  \frac{1}{2! 2!}
 \int_{x_1, x_2,x_3,x_4}   \left.
\frac{\delta^4 S_{eff}}{\delta \overline\psi (x_1) \delta  \psi(x_2) \delta \overline\psi (x_3) \delta  \psi(x_4)} 
 \right|_{\psi=\bar{\psi}=0}
 \overline\psi (x_1) \psi (x_2 )   \overline\psi (x_3) \psi (x_4)+
h.o. , 
\end{eqnarray}
where $\int_{x_1, x_2} = \int d x_1 \int d x_2$,   $h.o.$ stands for 
(even) higher order  derivatives and the 
 odd powers   must disappear since  they are 
 calculated for  $\psi, \overline\psi \to 0$ at the end.

The second order  term produces the following contribution:
\begin{eqnarray} \label{2nd-order-1}
S_{eff}^{(2)} =  g_4 \;  \int_x \, \mbox{tr} \,
\left[
 \left(    S_0 (x-y) \lambda_i \right) \; (\bpsi \lambda_i \psi)
+  \left(  S_0 (x-y) i \gamma_5 \lambda_i \right)  \; ( \bpsi i \gamma_5 \lambda_i \psi )
  \right],
\end{eqnarray}
where $S_0(x-y) = \left. S(x-y) \right|_{\overline{\psi},\psi \to 0}$
and tr stands for the traces of discrete indices.
The operatorial  coefficients of the quark bilinears 
will be resolved separately from the quark-antiquark 
 bilinears.
Those operators when resolved in momentum space will contribute
to the effective coupling constants.
This way, expression (\ref{2nd-order-1})  can  then be rewritten as:
\begin{eqnarray}
S_{eff}^{(2)} =  \int_x   \; tr \;  \left\{- g_4 \;   \lambda_{0}
 \; \left( \int \frac{d^4 k}{(2 \pi)^4} \, \frac{1}{   \gamma \cdot k - M^* } 
\right)\right\}
 \bar{\psi} (x)  \lambda_{0} \psi (x)  , 
\end{eqnarray}
where  the local limit was considered
and
 $tr \left[ \gamma_5 \right] = 0$, $tr \left[ \lambda_i\right] = 0 $ for $i\neq 0$.
The traces were  calculated   yielding one mass term for each of the 
quark flavors.
Therefore this second order term of the expansion produces  a correction to the quark masses
that can be written, in the Euclidean 
momentum space, as
\begin{eqnarray} \label{mass-corr}
\Delta m_f = 16 g_4 N_c \int \frac{d^4 k_E}{(2 \pi)^4} \, 
\frac{M^*_f}{k_E^2 + {M^{*}_f}^2}.
\end{eqnarray}
It is interesting to emphasize that this expression
is different from the effective mass
 given by the expression (\ref{Meff}) although
it was calculated in terms 
of the same parameters and  cutoff considered in the last section for the GAP equations.
It is worth to remember that  we adopt the model in which the NJL coupling constant
 is inversely proportional to the 
   gluon  effective mass,  expression
(\ref{g4}), and the resulting effective quark masses $m^*$ (and $M^*$) are  shown in Table I
 for different values of $M_G$.
Although the departing point was a $U(3)$ NJL model, from here on the calculations will be restricted to
the $SU(3)$ model.

The fourth order term in expression  (\ref{deriv-exp}) is calculated next
for zero momentum exchange.
The  
operators which are not contracted with the quark fields will be resolved 
yielding the coupling constant. This garantees the 
chiral invariance of the original interaction.
In the limit of zero momentum transfer this term can be written  as:
\begin{eqnarray}
S_{eff}^{(4)} = - 16 g_4^2 N_c \; tr \int_x
  \left[
\int \frac{d^4 k}{(2\pi)^4} \, 
\left( \frac{1}{\gamma \cdot k  - M^*}\right)^2 \, 
 \lambda_j^2 \right]
  \left\{
 (\overline{\psi} \lambda_i \psi)^2  
+ 
 (\overline{\psi} i \gamma_5 \lambda_i \psi)^2  \right\}
= \tilde{g}_4 \int_x  \left\{
 (\overline{\psi} \lambda_i \psi)^2  
+ 
 (\overline{\psi} i \gamma_5 \lambda_i \psi)^2  \right\} ,
\end{eqnarray}
where it was used that $\gamma_5^2 = I$ and $tr (\lambda_i \lambda_j) = 2  \delta_{i j}$.
This is a one-loop correction to the    NJL coupling constant which will be calculated
by regularizing the integral for  an  Euclidean  four momentum cutoff.
This coupling constant can be written as:
\begin{eqnarray} \label{g4-eff}
\tilde{g}_4 =  4 g_4^2 N_c \int \frac{d^4 k_E}{(2\pi)^4} \sum_{f}
\frac{ k^2_E- {M^*_f}^2}{(k^2_E + {M^*_{f}}^2)^2 }.
\end{eqnarray}
In Table I values for this coupling for different values of the effective gluon mass  from the GAP equation
are shown.

The sixth 
 order terms of the expansion will similarly be given
  by
\begin{eqnarray} \label{Seff-6}
S_{eff}^{(6)} &=& \frac{1}{3! 3!} \int_{x_{i=1,2...,6}}
\left.
\frac{\delta^6 S_{eff}}{
\delta \overline\psi (x_6) \delta  \psi(x_5)
\delta \overline\psi (x_4) \delta  \psi(x_3)
\delta \overline\psi (x_2) \delta  \psi(x_1)
} 
 \right|_{\psi=\overline{\psi}=0}
\overline\psi (x_6)  \psi (x_5)
\overline\psi (x_4)  \psi (x_3)
\overline\psi (x_2)  \psi (x_1).
\end{eqnarray}
After the factorization of the operators that yield the effective coupling constant, similarly to the previous terms,
 the following identity is used:
\begin{eqnarray}
tr (\lambda_i \lambda_j \lambda_k) =  D_{ijk},
\end{eqnarray}
being that in the $SU(3)$ case it reduces to 
$D_{ijk} = 2 \left( d_{ijk} + i f_{ijk} \right)$
where  $d_{ijk}$ and  $f_{ijk}$ are the symmetric and anti-symmetric $SU(3)$ tensors.
Expression (\ref{Seff-6})
 can then be written as:
\begin{eqnarray}
S_{eff}^{(6)} &=& 32  g_4^3 N_c \int_x \, \int \frac{d^4k}{(2\pi)^4} \, 
\sum_{f=u,d,s}
 \left( \frac{1}{  (\gamma \cdot k - M^*_{f})^3 }  \right)
\nonumber
\\
&& \times
 \left\{
 \frac{ d_{ijk}}{18} \left[  (\overline{\psi} \lambda_i \psi)  (\overline{\psi} \lambda_j \psi)  (\overline{\psi} \lambda_k \psi)
-  3(\overline{\psi} i \gamma_5 \lambda_i \psi)(\overline{\psi} i \gamma_5 \lambda_j \psi) 
(\overline{\psi} \lambda_k \psi) \right]   \right\},
\end{eqnarray}
where the antisymmetric component is zero.
This term has exactly the flavor structure  of the $SU(3)$ deteminantal t' Hooft  interaction \cite{NJL2,NJL3,NJL4} 
that can be written as:
\begin{eqnarray} \label{thooft-eff}
S_{eff}^{(6)} &=& \tilde{g}_6 \int_x \;  
 \frac{ d_{ijk}}{18} \left[  (\overline{\psi} \lambda_i \psi)  (\overline{\psi} \lambda_j \psi)  (\overline{\psi} \lambda_k \psi)
-  3(\overline{\psi} i \gamma_5 \lambda_i \psi)(\overline{\psi} i \gamma_5 \lambda_j \psi) 
(\overline{\psi} \lambda_k \psi) \right],
\end{eqnarray}
where the
 effective coupling, with the same  covariant   Euclidean momentum cutoff ,  is given by:
\begin{eqnarray}
\tilde{g}_6 &=&  -  32 \; g_4^3 N_c \int \frac{d^4 k_E}{(2\pi)^4}
\;  
\sum_{f=s,d,u}  \;
\frac{   3k^2_E  M_f^* - {M_f^*}^3 }{
(k^2_E + {M^*_f}^2)^3 }.
\end{eqnarray}
This coupling constant is related to the usual definition of the 't Hooft term
 ($\kappa$ in Refs. \cite{instability,stability-8th,NJL2,NJL3}) 
by $\tilde{g_6} = \frac{9}{16} \kappa$.
Numerical values for  this coupling constant,  with the same  parameters as before, are  shown 
in Table I . They are all negative in agreement with phenomenological values.

\begin{table}[ht]
\caption{ 
Values of the mass correction calculated with the parameters fitted  from the GAP equations (\ref{gaps-phi-sigma}) - $M_G$, $\Lambda$ 
and quark current masses $m_u = 3$ MeV $m_d= 6.6$ MeV, $m_s= 90.6$ MeV, and gauge coupling $g$ from Ref. \cite{lattice}, considering the NJL coupling constant ($g_4$) obtained from the effective model.} 
\centering 
\begin{tabular}{c c c c c c c c c c c c} 
\hline\hline 
$M_G$ [$\Lambda$] & $\Delta m_u$  [$M^*_u$]& $\Delta m_d$  [$M^*_d$]   &  $\Delta m_s$ [$M^*_s$]    & $g_{4}$  
  & $\tilde{g}_4$  & $\tilde{g}_6$ &
$g_1^{(8)}$ & $g_2^{(8)}$ 
\\
 (MeV)  &  (MeV)&   (MeV) &   (MeV)   &  (GeV$^{-2}$) 
  &  (GeV$^{-2}$) & (GeV$^{-5}$) &
(GeV$^{-8}$) & (GeV$^{-8}$) 
\\
\hline
\\ [0.5ex]
600 [651] & 271 [274] & 273 [281] & 303  [395] & 12 &  2.8   & - 1100 &  6526   & 1957
\\ 
650 [706] & 294 [297] & 296 [303] & 328 [419]  & 10  &  2.5  &  - 748   &  3469  & 1041
\\ 
700 [760] & 312 [319] & 319 [326] & 352 [352] &  8   &  2.2   &    -  520  &  1930  & 579
\\ 
800 [870] & 359 [365] & 366 [372] & 401 [492] & 6  &  1.7    & -  278  &  670  & 201
\\[1ex] 
\hline 
\end{tabular}
\label{table:results-2} 
\end{table}

The eighth order term is calculated from the following derivative:
\begin{eqnarray}
\left.
\int_{x_{i=1...8}} \frac{\delta^8 S_{eff}}{
\delta \overline\psi (x_8) \delta  \psi(x_7)
\delta \overline\psi (x_6) \delta  \psi(x_5)
\delta \overline\psi (x_4) \delta  \psi(x_3)
\delta \overline\psi (x_2) \delta  \psi(x_1)} 
 \right|_{\psi=\overline{\psi}=0}
 \overline\psi (x_8)  \psi (x_7)
 \overline\psi (x_6)  \psi (x_5)
 \overline\psi (x_4)  \psi (x_3)
 \overline\psi (x_2) \psi (x_1).
\end{eqnarray}
This term  will be also resolved in the limit of zero   momentum exchange.
With the same factorization of the traces, it 
can be written as:
\begin{eqnarray} \label{8th-1}
S_{eff}^{(8)} &=& - 16 \frac{(2 g_4)^4 N_c }{4 ! 4 !} \; \; \int d^4 x \,\; \int_{x_{i=1...8}}
\; tr \left[ \lambda_i \lambda_j\lambda_k \lambda_l \right]  \;\;  
\left( \sum_f \frac{1}{ (i \slashed{\partial} - M^*_{f} )^4 } \right) 
\nonumber
\\
&& \times
\sum_{a\neq b\neq c \neq d=1,3,5,7 ; e\neq f\neq g\neq h=2,4,6,8} 
\left[  (\overline{\psi} N_{a e}^i   \psi) (\overline{\psi} N_{b f}^j   \psi) (\overline{\psi} N_{c g}^k   \psi)
 (\overline{\psi} N_{d h}^l   \psi)
\right]  ,
\end{eqnarray}
where different combinations of the operators
$N_{ae}^i \equiv (\lambda_i + i \gamma_5 \lambda_i)_{ae}$ were found and are
they are defined below.
 The traces over $(\gamma_5)^{2n}$ ($n$ integer) were  calculated for these coefficients.
Only two types of terms in expression (\ref{8th-1}) are nonzero, namely those for which 
the flavor structure of the quark bilinears arranges  in the following forms:
\begin{eqnarray} \label{structures-1}
K_1 &\to& tr (N_{14}N_{32})(N_{56}N_{78})
\\
\label{structures-2}
K_2 &\to& tr (N_{16}N_{74} N_{52}N_{38}).
\end{eqnarray}
All  the other terms  either 
disappear, since $tr \left[ \gamma_5 \right] = tr \left[ \lambda_i \right] = tr \left[ \gamma_5^3 \right] =0$,
 or they reduce to one of these two terms.

To rewrite  expression (\ref{8th-1})
the following $SU(3)$ relations were used:
\begin{eqnarray} \label{su3-relat}
 tr \left( \lambda_i \lambda_j \lambda_k \lambda_l \right) &=& 
16 \left(
\frac{1}{12} \delta^{ij} \delta^{kl} + \frac{1}{8} h^{ija} h^{akl}
\right), \;\; \mbox{for} \;\; h_{ija} = d_{ija} + i f_{ija},
\nonumber
\\
h_{ija} h_{akl} &=& d_{ija} d_{akl} - f_{ija} f_{akl}
+ i \left(  d_{ija} f_{akl} + f_{ija} d_{akl}
\right),
\nonumber
\\
tr \lambda_i \lambda_j &=& 2 \delta_{ij},
\end{eqnarray}
as well as the $SU(3)$ Jacobi identity and the (anti)symmetry 
of the tensor $d_{ija}$ ($f_{ija}$).
By considering a simplified notation with  $s_i = \overline{\psi} \lambda_i \psi$ and
$p_i = \overline{\psi} \lambda_i  \gamma_5 \psi$, 
the first structure, expression (\ref{structures-1}), can be written as:
\begin{eqnarray}
K_1 \to \frac{16}{12} \left[ 
s_i ^2+  p_i^2  
\right]^2,
\end{eqnarray}
where we used that $h_{iia}=0$
and  $tr \left[ \gamma_5 \right] = tr \left[ \gamma_5^3 \right] = 0$.

For the second term it follows:
\begin{eqnarray}
K_2 \to 4 
\frac{16}{12} \left( s_i^2 + p_i^2 \right)^2
+  \frac{16}{8}
\left[
d_{ija} d_{akl} \left( s_i s_j s_k s_l + p_i p_j p_k p_l + 2 s_i s_j p_k p_l \right)
- 
4 f_{ija} f_{akl}  s_i p_j s_k p_l 
\right]
\end{eqnarray}
Now the chiral projectors,  
$P_{R,L} = \frac{1}{2} \left( 1 \pm \gamma_5 \right)$, 
 can be used to rewrite the 
terms above.
By resolving all the traces (Dirac, flavor and color)
 of the corresponding coefficients,  back to expression (\ref{8th-1}), it yields:
\begin{eqnarray} \label{8th-2}
S_{eff}^{(8)} &=& \;  \tilde{g}_8 \;   \int_x \, 
 \left\{ \left( \frac{16}{12}  +  4  \frac{16}{12}  \right)
\left(
 \bar{\psi}  P_R \psi\; \bar{\psi} P_L \psi\; 
\right)^2  
+ \frac{16}{8} \left(
 \bar{\psi}  P_R \psi\; \bar{\psi} P_L \psi\; \bar{\psi} P_R  
\psi\; \bar{\psi} P_L \psi
\right) 
 \right\},
\end{eqnarray}
where  the following effective coupling constant  was defined, using  Euclidean momenta
with the same cutoff as before:
\begin{eqnarray}
\tilde{g}_8 =   16 \times 16\; \frac{(2 g_4)^4 N_c}{ \; 4 ! \; 4 !}
\int \frac{d^4 k_E}{(2\pi)^4} \, 
\left(  \sum_{f} \frac{  k^4_E - 6 k^2_E   {M_{f}^*}^2 + {M_{f}^*}^4
}{ (k^2_E + { M^*_{f}}^2 )^2
 ( k^2_E + {M^*_{f}}^2)^2} \right).
\end{eqnarray}
The two terms in expression (\ref{8th-2})  are precisely the most general 
$SU(3)$ chiral  invariant Lagrangian interactions considered 
by Osipov {\it et al} \cite{stability-8th,stability-8th-2} which can be rewritten as:
\begin{eqnarray} \label{V8q} 
{\cal L}_{eff,8} = g_{2}^{(8)}
\left(
 \bar{\psi}  P_R \psi\; \bar{\psi} P_L \psi\; \bar{\psi} P_R  
\psi\; \bar{\psi} P_L \psi
\right) +
 g_{1}^{(8)}
\left(
 \bar{\psi}  P_R \psi\; \bar{\psi} P_L \psi\; 
\right)^2  .
\end{eqnarray}
The  couplings $g_1^{(8)} = \tilde{g}_8   \frac{80}{12} $
and $ {g}_2^{(8)} = \tilde{g}_8  \frac{16}{8} $ have different relative weights
in expression (\ref{8th-2})  which    behave in 
the way as suggested  in  Refs. \cite{stability-8th,stability-8th-2} for the stability 
of the ground state, i.e. $g_1^{(8)} > g_2^{(8)}$.
In Table I,  some values for the effective  coupling constants  $g_1^{(8)}$ and  $g_2^{(8)}$ are 
given as functions 
of the same values for free parameters and gluon effective mass (or cutoff) 
from  the GAP equations.

The values of  the masses and effective coupling constants shown in Table I are 
in very good agreement with phenomenological fits in the investigation of 
the spontaneously broken chiral symmetry and  light hadron structure 
\cite{NJL2,stability-8th,stability-8th-2}.
The   effective quark masses $m_f^*$
 are very close to the 
values for the effective masses $M^*_f$ obtained  from the scalar quark condensates $S_f$,
but a little smaller for the strange quark mass.
This seems to justify the use of $M^*_f$ as equivalent to $m^*_f$.
The NJL effective coupling constant $g_4$ and $\tilde{g}_4$ are of the order of the usual
coupling considered in different versions of the model \cite{NJL2,NJL3,stability-8th,stability-8th-2}.
The sixth order coupling constant is slightly smaller than  the phenomenological fits for the  't Hooft coupling  
($\kappa \simeq - 770 \to - 1100$ GeV$^{-5}$) as it was considered in Refs.
\cite{instability,stability-8th,stability-8th-2}, since $\kappa = \frac{16}{9} \tilde{g}_6$.
Finally the eight order terms are also slightly higher  than  the values considered in the 
phenomenolgical fits by Osipov and collaborators
\cite{instability,stability-8th}, following nearly the systematics for the two different coupling constants,
i.e. $g_1 > g_2$ what is required by the vacuum stability conditions analysis 
found in those references. Both  $g_1^{(8)}$ and $g_2^{(8)}$ are positive, although they
compare well with 
the phenomenological fits in the ranges 
of  $g_1 \sim 1000 \to 6000$ GeV$^{-8}$ and $g_2 \sim -130 \to 320$ GeV$^{-8}$.

\section{ Summary and Conclusions}

In this work,  low energy effective quark interactions  were derived 
by considering vacuum polarization   for 
a QCD-based Nambu-Jona-Lasinio model, 
where  the NJL coupling is proportional to 
the zero momentum QCD coupling constant $g^2$ and inversely 
proportional to the effective gluon mass \cite{barros-braghin}.
The quark field was 
separated into two components, one that condenses into scalar quark-antiquark condensate
and another one corresponding to interacting  quarks, the 
quasiparticles of the model.
By integrating out the first component  with the  introduction of the 
usual scalar/pseudoscalar variables, 
 an effective model for the interacting quarks  in terms of the 
vacuum values of three auxiliary variables $\phi^0, S_i^{0}, P_i^{0}$ was obtained.
Since these variables can be associated to typical condensates 
 of the QCD vacuum, their
 gap equations were calculated and solved
by extremizing the effective potential at zero quark field and by considering 
 an unique covariant Euclidean cutoff as usually done.
The resulting quark and gluon effective masses, $M^*_f$ and  $M_G$, are basically the same as
 those
obtained in different approaches.
The quark determinant 
was expanded in powers of bilinears $\overline{\psi} \Gamma \psi$
yielding corrections to the interacting quark (quasiparticles) masses 
and effective multiquark couplings.
The resulting quark effective  coupling constants
were calculated by factorizing each term of the determinant expansion  
following nearly the lines of  the lowest order derivative expansion
yielding chiral invariant multiquark interactions in terms of 
 the same covariant cutoff
fixed in the solution of the GAP equations, being 
 the model  non renormalizable.
Besides that, if in one hand this calculation might
be seen as a first analysis to be supplemented by a renormalization group investigation,
in another hand, the values for the masses and effective coupling constants shown in Table I are 
in very good agreement with phenomenological fits in the investigation of 
the spontaneously broken chiral symmetry and  light hadron structure \cite{NJL2,stability-8th}.
The vacuum polarization corrections for the effective interacting quark masses ($m_f^*$)
 are very close to the 
values for the effective masses $M^*_f$  from the scalar quark condensates $S_{0,f}$,
but a little smaller for the strange quark mass.
This seems to allow the identification of both masses $m_f^* \sim M_f^*$.
The correction to the NJL  coupling constant  $\tilde{g}_4$ is smaller than the  usual
coupling considered in different versions of the model but of the same order of magnitude
 \cite{NJL2,stability-8th}.
The sixth order coupling constant has  slightly smaller modulus than the phenomenological fits for the
  't Hooft coupling   as it was considered in Refs.
\cite{instability,stability-8th}, 
$\kappa \simeq - 770 \to - 1100$ GeV$^{-5}$,
by identifying  $\kappa = \frac{16}{9} \tilde{g}_6$.
Finally the eighth order term is in good agreement with the values considered in the 
phenomenological fits by Osipov and collaborators
\cite{instability,stability-8th}, following nearly the systematics for the two different coupling constants,
i.e. $g_1 > g_2$ what is required by the vacuum stability conditions analysis 
found in those references.
They have values  $g_1 \sim 1000 \to 6000$ GeV$^{-8}$ and $g_2 \sim -130 \to 320$ GeV$^{-8}$.
It is interesting to note that  some of the  resulting effective quark interactions
correspond the  ones obtained from
 other derivations based on instanton physics, in particular the 
 sixth order interactions and seemingly and the eighth order one \cite{simonov,thooft}.
Although these two different calculations might provide  a sort of  double counting
of  QCD effects, i.e.     from  instanton physics and  QCD condensates, 
 the discussion of this  issue is outside the scope of the present work.
The approach considered in this work is  a systematic framework which 
allows for  improvements and to account further effects, in particular from 
the QCD vacuum.
Therefore it   might  be valuable
for  the analysis of the stability of the  QCD expansion in quark currents.
Moreover, it  allows for computing corrections to different   effective quark interactions
 other  than those found here, 
such as derivative couplings neglected above.
This program
would help to pin down   the correct physical value for the effective couplings at the 
desired energy scale.

\section*{Acknowledgements}

The authors thank short conversation with G.I. Krein.
F.L.B acknowledges financial support from CNPq.

\end{document}